\documentclass{article}

\usepackage{graphicx}     
\usepackage{amsmath,amssymb} 
\usepackage{booktabs}      
\usepackage{caption}      
\usepackage{enumitem}      
\usepackage{array,tabularx}
\usepackage{xcolor}       
\usepackage{hyperref}       
\usepackage{natbib}
\usepackage{longtable}
\usepackage{supertabular}
\usepackage{newtxtext,newtxmath}
\usepackage[T1]{fontenc}  \usepackage{ae,aecompl}
\usepackage{graphicx} \usepackage{amsmath}     \usepackage{amssymb}
\usepackage{epsfig}  \usepackage{lineno}  \usepackage{overpic}
\usepackage{hyperref}

\def\apj{ApJ}
\def\apjl{ApJL}
\def\apjs{ApJS}
\def\mnras{MNRAS}
\def\araa{ARA\&A}
\def\aap{A\&A}
\def\aj{AJ}
\begin{document}

\begin{center}
    { \large \bfseries Revisiting the Origin of the Star-Forming Main Sequence Based on a Volume-Limited Sample of $\sim$25,000 Galaxies}

    \vspace{0.3cm}

    {\large
    Yang Gao$^{1,2,\dagger}$, Shujiao Liang$^{1}$, Qinghua Tan$^{2}$, Enci Wang$^{3}$, Huilan Liu$^{1}$, Hongmei Wang$^{1}$, Tao Jing$^{4}$, Xiaolong Wang$^{5,6,7}$, Kaihui Liu$^{1}$, Ning Gai$^{1}$, Yanke Tang$^{1}$, Yifan Wang$^{1}$, Yutong Li$^{1}$
    }

    \vspace{0.3cm}

     \footnotesize
    $^1$ Shandong Key Laboratory of Space Environment and Exploration Technology, College of Physics and Electronic Information, Dezhou University, Dezhou 253023, China $^{\dagger}$ Corresponding author: \texttt{gao14681@mail.ustc.edu.cn}\\
    $^2$ Purple Mountain Observatory, Chinese Academy of Sciences, 10 Yuanhua Road, Nanjing 210023, China\\
    $^3$ CAS Key Laboratory for Research in Galaxies and Cosmology, Department of Astronomy, University of Science and Technology of China, Hefei 230026, China\\
    $^4$ Department of Astronomy, Tsinghua University, Beijing 100084, China\\
    $^5$ College of Physics, Hebei Normal University, 20 South Erhuan Road, Shijiazhuang 050024, China\\
    $^6$ Guo Shoujing Institute for Astronomy, Hebei Normal University, Shijiazhuang 050024, China\\
    $^7$ Hebei Advanced Thin Films Laboratory, Hebei Normal University, Shijiazhuang 050024, China
    \normalsize
    \vspace{0cm}
    
\end{center}

\begin{abstract}
We revisit the extensively debated star-forming main sequence (SFMS)—a tight correlation between the star formation rate and stellar mass in both kiloparsec-resolved and integrated galaxies.
We statistically explore the fundamental drivers of star formation at global scales, using a large volume-limited sample of 24,954 local star-forming galaxies to overcome the limitations of previous works. 
Based on the mid-infrared 12 $\mu$m luminosity, stellar mass, and $g-r$ color, we estimate the molecular gas mass for the considered sample. 
At galaxy-wide scales, we establish global relations between the surface densities of the star formation rate ($\Sigma_{\rm SFR}$), stellar mass ($\Sigma_{\ast}$), and molecular gas mass ($\Sigma_{\rm mol}$). These global density relations are connected with and follow similar trends as the resolved SFMS, the Kennicutt–Schmidt (KS) relation, and the molecular gas main sequence (MGMS).
Taking advantage of this large catalog, we show that the scatters in the global KS and MGMS relations are smaller than that of the global relation between $\Sigma_{\rm SFR}$ and $\Sigma_{\ast}$, and their Pearson correlation coefficients are higher. More importantly, multivariate regression and partial correlation analyses demonstrate that the apparent $\Sigma_{\rm SFR}$-$\Sigma_{\ast}$ correlation is entirely mediated by $\Sigma_{\rm mol}$, with its best-fit parameters directly derivable from those of the KS and MGMS relations.
Overall, our findings suggest that the correlation between stellar mass and molecular gas, as well as that between molecular gas and star formation, are more direct and fundamental. The star-forming main sequence thus appears to be a natural by-product of these two tighter relations.
\end{abstract}


\section{Introduction}
Most star-forming (SF) galaxies are observed to be located in a narrow sequence in the stellar mass ($M_\ast$) vs. star formation rate (SFR) plane, which has been called the star-forming main sequence (SFMS) \citep{Brinchmann2004,2007ApJ...670..156D,2007ApJS..173..267S}. 
This tight relation not only serves as a key diagnostic for galactic properties but also provides a well-established empirical foundation for modern galaxy evolution studies. It forms the basis for classifying galaxies as star-forming, starburst, or quenched based on their position relative to the sequence.
 Although this relation has been observed at kpc scales as the resolved star-forming main sequence (rSFMS) \citep{Cano2016}, the physical connection between the instantaneous star formation rate and the total assembled stellar mass remains less well understood.

 Within the framework of galactic evolution, stars form as gas accumulates under gravity and collapses \citep{Shi2011,Schinnerer2024}. As described by the "gas-regulator" model \citep{2013ApJ...772..119L}, star formation is causally determined by the abundance of molecular gas, following the Kennicutt--Schmidt (KS) law ($\Sigma_{\rm SFR}$ vs. $\Sigma_{\rm mol}$) established via direct CO observations.

  The COLD GASS survey demonstrated that a galaxy's position in the SFR–$M_\ast$ plane is primarily determined by its gas content and star formation efficiency \citep{Saintonge2011b,Saintonge2012,Saintonge2016}. This gas-centric framework has been reinforced through resolved studies; analysis of ALMA and MaNGA data showed that the correlation between $\Sigma_{\rm SFR}$ and $\Sigma_{\ast}$ becomes insignificant when controlling for $\Sigma_{\rm mol}$ \citep{Lin2019,Baker2022MNRAS.510.3622B}. Consequently, the observed rSFMS can be understood as emerging from the combination of the fundamental KS law and the Molecular Gas Main Sequence (MGMS, $\Sigma_{\rm mol}$ vs. $\Sigma_\ast$) \citep{Lin2019,wang2019,Ellison2020MNRAS.493L..39E}, with the latter potentially regulated by the local gravitational potential \citep{Wong2013ApJ...777L...4W} or the consumption of molecular {gas} due to  $\Sigma_\ast$-dependent gravitational instabilities \citep{Zheng2013}. 
  
This gas-driven paradigm directly challenges alternative interpretations. Some studies have argued for the primacy of stellar mass, finding a stronger rSFMS correlation or identifying total baryonic mass as the key regulator, consistent with self-regulation models \citep{2021ApJ...909..131B,Sanchez2020,Ostriker2010}. Furthermore, a unifying perspective from the EDGE-CALIFA survey suggests that $\Sigma_{\rm SFR}$, $\Sigma_{\ast}$, and $\Sigma_{\rm mol}$ follow a line (or cylinder) in 3D space, implying that 2D relations are merely projections without a primary driver \citep{2021MNRAS.503.1615S}.

 Resolving these conflicting interpretations—regarding gas as the direct fuel, mass as a global regulator, or a higher-dimensional manifold—is essential for identifying the principal mechanism regulating star formation in galaxy disks. 
 However, previous studies based on different galaxy samples have yielded inconsistent results, preventing a coherent picture from emerging. This lack of consensus likely stems from the limitations of those samples; for instance, small galaxy numbers (tens or hundreds), constituting a narrow range in galaxy properties. 

  To overcome these limitations, we employ a large and volume-limited sample of 24,954 star-forming galaxies in this study, which provides a clear empirical basis for distinguishing between competing physical scenarios. We extend the insights from resolved studies to galaxy-integrated properties by directly comparing the intrinsic strength and scatter between the three scaling relations of galaxy-wide average surface densities. Hereafter, we refer to these relations measured on galaxy-wide scales as the global surface density relations, which correspond to the global intensive relations defined by \citet{2021MNRAS.503.1615S}. Through identifying which of these relations exhibits the strongest correlation and smallest scatter, we directly and statistically test whether the SFMS is a fundamental relation or an emergent byproduct of more fundamental gas–star formation connections. To further disentangle the direct drivers of star formation and address potential dependencies among these measurements, we also employ partial correlation analysis and multivariate regression in this work. 

 The remainder of this paper is structured as follows. The next section introduces the sample and data used. The molecular gas fraction, depletion time-scale, and the global surface density relations between $\Sigma_{\rm SFR}$, $\Sigma_{\ast}$, and $\Sigma_{\rm mol}$ are detailed in Section~\ref{sec:results}. Section~\ref{sec:Discussion} connects the global relations with resolved ones and discusses whether the global relation between $\Sigma_{\rm SFR}$ and  $\Sigma_{\ast}$ is less fundamental than the two others.
 Finally, we summarize our findings in Section~\ref{sec:summary}. Throughout the paper, distance-dependent quantities are calculated  by assuming a
standard flat ${\Lambda}$CDM cosmology with a matter density parameter
$\Omega_{\rm m}=0.3$, a dark energy density parameter
$\Omega_\Lambda=0.7$, and a Hubble constant of $h$ = 0.7.
\section{Sample and Statistical Properties}
\subsection{SDSS Volume-limited MS Sample }
To obtain a representative sample of local main-sequence (MS) galaxies, we constructed a parent sample consisting of 40,974 galaxies, forming a volume-limited sample with redshifts $z<0.05$ and stellar masses $M_\ast>10^{9.5}$ M$_\odot$, which was cross-matched from the NASA Sloan Atlas Catalog (NSA) \citep{Blanton2011}, AllWISE Source Catalog \citep{Wright2010}, and \textit{GALEX}-SDSS-\textit{WISE} Legacy Catalog (GSWLC) \citep{Salim2016,Salim2018}. 
From this parent sample, main-sequence galaxies were selected along the curved star-forming main sequence defined by \citet{Saintonge2016}:
log~$(SFR/[{\rm M_\odot}/yr]) = -2.332 x + 0.4156 x^2  -0.01828 x^3$, where x = log~$({M_\ast}/{\rm M_\odot})$.
Galaxies within a vertical deviation of 0.4 dex from this relation and with signal-to-noise ratio S/N > 2 for all relevant parameters were retained, yielding a final sample of 24,954 galaxies. This selection threshold is consistent with previous works \citep{Saintonge2017} and ensures a representative sample of star-forming systems while maintaining a manageable sample size. The resulting sample has a scatter of $\Delta(MS)$ = 0.11 dex.

\subsection{Galaxy Properties}
\label{Properties}
Stellar mass and SFR, along with their uncertainties, were obtained from GSWLC for all galaxies, which are derived from state-of-the-art UV/optical SED fitting with the CIGALE code incorporating GALEX FUV/NUV and SDSS ugriz photometry using a modified dust attenuation law and emission line corrections. This approach provides more reliable estimates for low-SFR galaxies, making it particularly robust for studies of local star-forming galaxy populations.

Other ultraviolet and optical parameters, including color index and light radius, were retrieved from 'nsa\_v1\_0\_1.fits'\footnote{http://nsatlas.org} \citep{Blanton2011}.  For instance, the NUV--$r$ color---a reliable indicator of the star formation status---is computed as the difference between the $K$-corrected NUV and $r$-band absolute magnitudes, both measured using an elliptical Petrosian model. 
To compare the global surface density relations with local/resolved relations between $\Sigma_{\rm SFR}$, $\Sigma_{\ast}$, and $\Sigma_{\rm mol}$, we adopted the parameter $R_{50}$ (which denotes the effective radius enclosing half of the elliptical Petrosian flux in the $r$ band) to compute the surface densities of stellar mass, SFR, and molecular gas mass following the method of \citet{Sanchez2020,2021MNRAS.503.1615S} (in brief, these global properties are divided by the area within 2 $R_{50}$; i.e., 4$\pi R^2_{50}$). This calculation is based on the assumption that stars, molecular gas, and SFR share the same spatial distribution. This assumption may be reasonable for $\Sigma_{\ast}$ and $\Sigma_{\rm mol}$, as noted in several previous studies which have observed a correspondence between CO and stellar radial profiles \citep{Leroy2008,Leroy2009,Saintonge2012}; however, applying it to $\Sigma_{\rm SFR}$ likely leads to underestimation of uncertainties, as discussed in Section~\ref{sec:Discuss1}.

\subsection{Estimated molecular gas masses}
After excluding galaxies whose \textit{WISE} 12 $\mu$m photometry is adversely affected, we combine $L_{12 \mu\textrm{m}}$, $M_\ast$, and $g-r$ to estimate their molecular gas masses ($M_{\rm mol}$) using the three-parameter estimator proposed by \citet{Gao2019}: 
log~$(M_{\rm mol}/[{\rm M_\odot}]) = 0.76  \log ({L_{12\ \mu\textrm{m}}}/[{\rm L_\odot}]) + 0.29  (g-r) + 0.29  \log ({M_\ast}/[{\rm M_\odot}]) - 0.77$.
This estimator is physically motivated by the use of color and stellar mass to statistically purify the correlation between CO-traced molecular gas and the PAH or dust emission probed by \textit{WISE} 12 $\mu$m, which spans a broad mid-infrared wavelength range \citep{Jarrett2011, Gao2022}. After calibration over a broad galaxy population, the estimator is expected to be most reliable for galaxies similar to those in our sample: $M_\ast > 10^{9.5}{\rm M}_\odot$ and $\log({\rm sSFR/yr^{-1}}) > -11.5$. However, possible deviations may occur for extreme populations such as starbursts or passive galaxies \citep{Gao2025}, which may represent a small fraction of our star-forming main-sequence sample.

All molecular gas masses in this work were converted to include the presence of heavy elements (mainly helium) by assuming a Galactic $\alpha_{\rm CO}$ = 4.3  M$_\odot$/[K km s$^{-1} {\rm pc^{2}}$], corresponding
  to $X_{\rm CO} = 2\times 10^{20}$ cm$^{-2}$/[K km s$^{-1}]$.

\subsection{Ancillary CO sample}
To characterize the population of this volume-limited sample, we compared it with the 198 CO-detected MS galaxies selected using the same criterion from the extended CO Legacy Data base for the GASS (xCOLD GASS) \citep{Saintonge2017}. The xCOLD GASS sample provides CO measurements for 532 mass-selected local ($0.01<z<0.05$) galaxies, which are representative of a galaxy population with $M_\ast>10^{9}$ M$_\odot$.

\section{Results}
\label{sec:results}

\subsection{Gas Fraction and Depletion Time Scaling Relations}

\label{sub_result1}
\begin{figure}[htbp]
\includegraphics[width=14.0 cm]{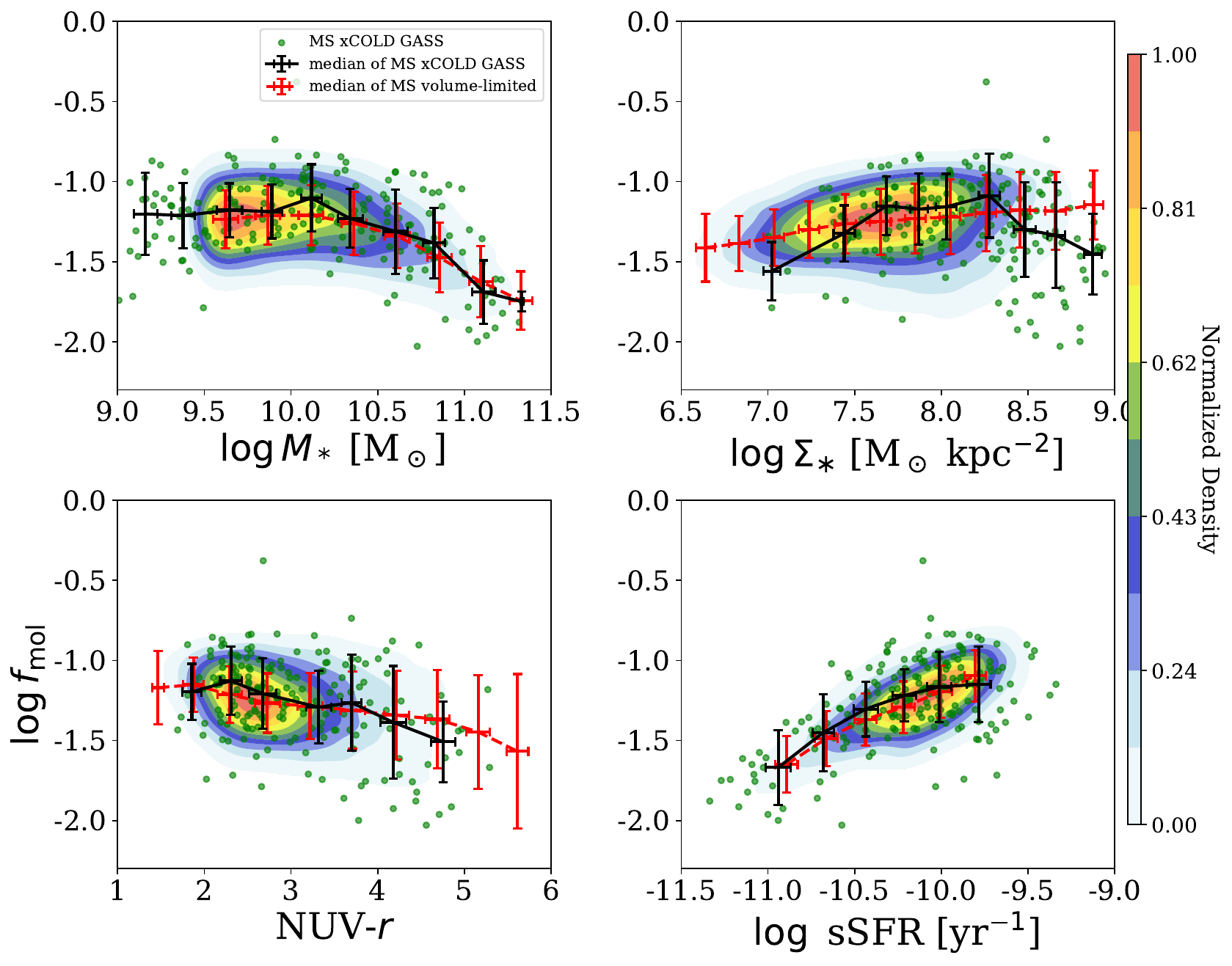}
\caption{
The molecular gas mass fraction ($f_{\rm mol} \equiv M_{\rm mol}/M_\ast$) is plotted against (from left to right): stellar mass ($M_\ast$), stellar mass surface density ($\Sigma_{\ast}$), NUV--r color, and the specific star formation rate (sSFR).  The green circles represent xCOLD GASS main sequence galaxies with CO detections, while the contour shows the distribution of our estimated volume-limited main sequence sample. The black and red points with error bars respectively indicate the median and 1 $\sigma$ scatters of the observed and estimated MS galaxies in each bin. \label{fig1}}
\end{figure}

\begin{figure}[htbp]
\includegraphics[width=14.0 cm]{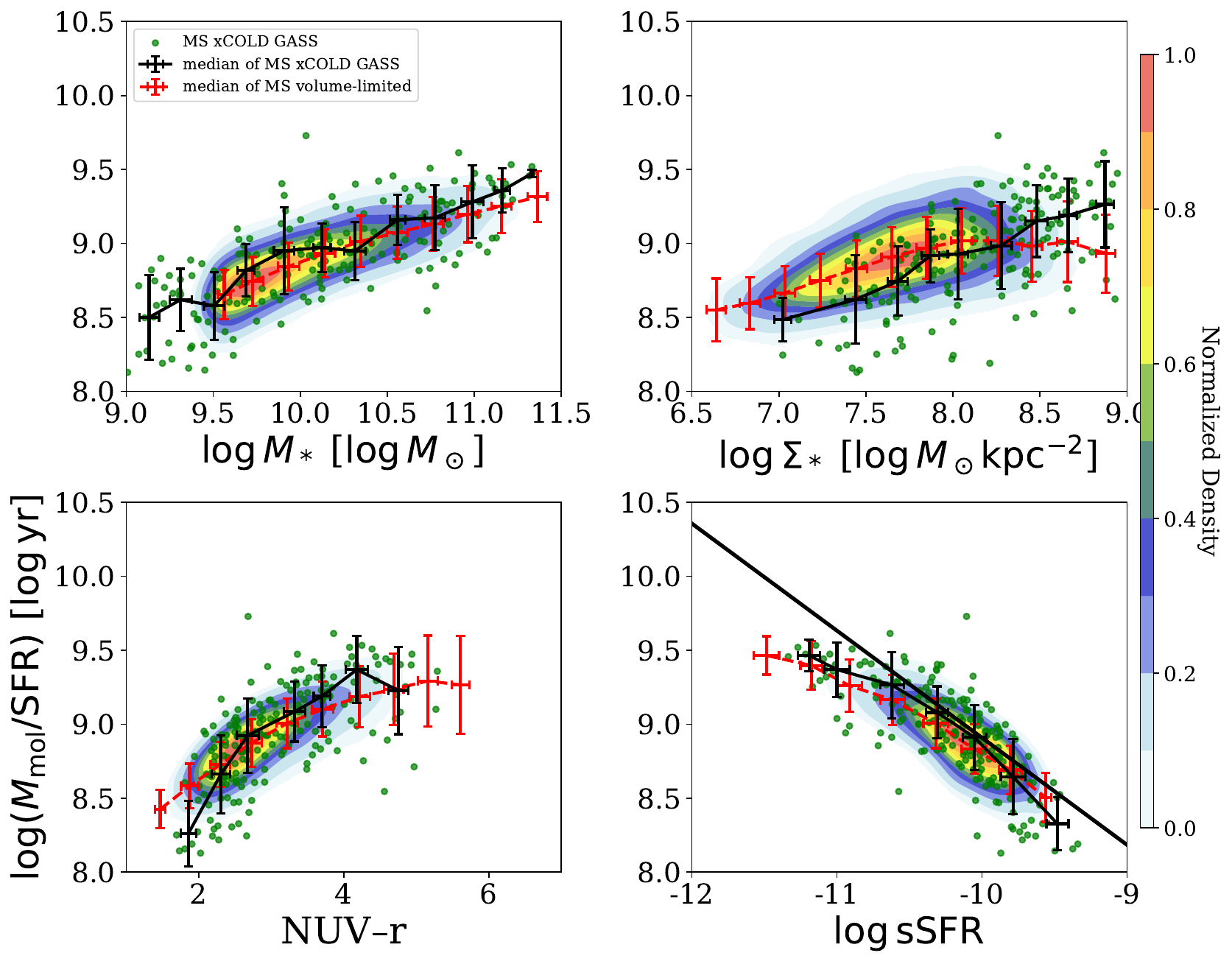}
\caption{Similar to Figure~\ref{fig1}, but for molecular gas depletion time scaling $t_{\rm dep}$(mol) relations. In the rightmost panel, we plot the adjusted best--fitting bisector linear relation provided by \citet{Saintonge2011b} based on the COLD GASS sample, assuming a constant Galactic conversion factor. \label{fig2}}
\end{figure}

The relationships between the molecular gas fraction ($f_{\rm mol}$) or its depletion time scale ($t_{\rm dep}$) and integrated galaxy properties are quantified in Figures~\ref{fig1}~and~\ref{fig2}.
A key finding is the strong agreement between the directly observed xCOLD GASS MS (green circles) and our estimated volume-limited MS sample (color contour and red median points). This consistency serves as a necessary check, showing that our large, statistically defined sample recovers the well-established scaling relations derived from direct CO observations \citep{Saintonge2017}. We note that our most primary conclusions regarding the global surface density relations in Figure~\ref{fig3} remained robust when using an alternative molecular gas mass estimator \citep{Gao2019} based solely on \textit{WISE} 12 $\mu$m luminosity. 

Notably, although all galaxies lie on the SFMS, both $f_{\rm mol}$ and $t_{\rm dep}$ exhibit clear correlations with sSFR across the sample. These trends indicate that galaxies with bluer colors and more active star formation tend to possess higher molecular gas fractions and shorter depletion timescales, which is consistent with general expectations for star-forming systems. 
One minor difference is that the slope derived from the estimated sample is somewhat shallower than that from the direct detections, although it remains consistent with the observed relations obtained using the metallicity-dependent $\alpha_{\rm CO}$ \citep{Saintonge2017}. This discrepancy may partly arise from the limitations of the estimator with respect to  some galaxies with low stellar mass or density, and/or the inclusion of more low-density galaxies where CO-dark gas is prevalent in the estimated complete sample \citep{Gao2025}. 

In constructing this statistically representative sample, we greatly increase the effective number of galaxies available for analysis and extend the explored parameter space beyond the limits of the direct CO-detected sample. This expansion enables a more robust characterization of the global relations presented in its intensive form and their scatter.

\subsection{The Scaling Relations between Surface Densities of Star Formation Rate, Stellar Mass, and Molecular Gas Mass }
\label{sub_result2}

\begin{figure}[htbp]
\includegraphics[width=14.0 cm]{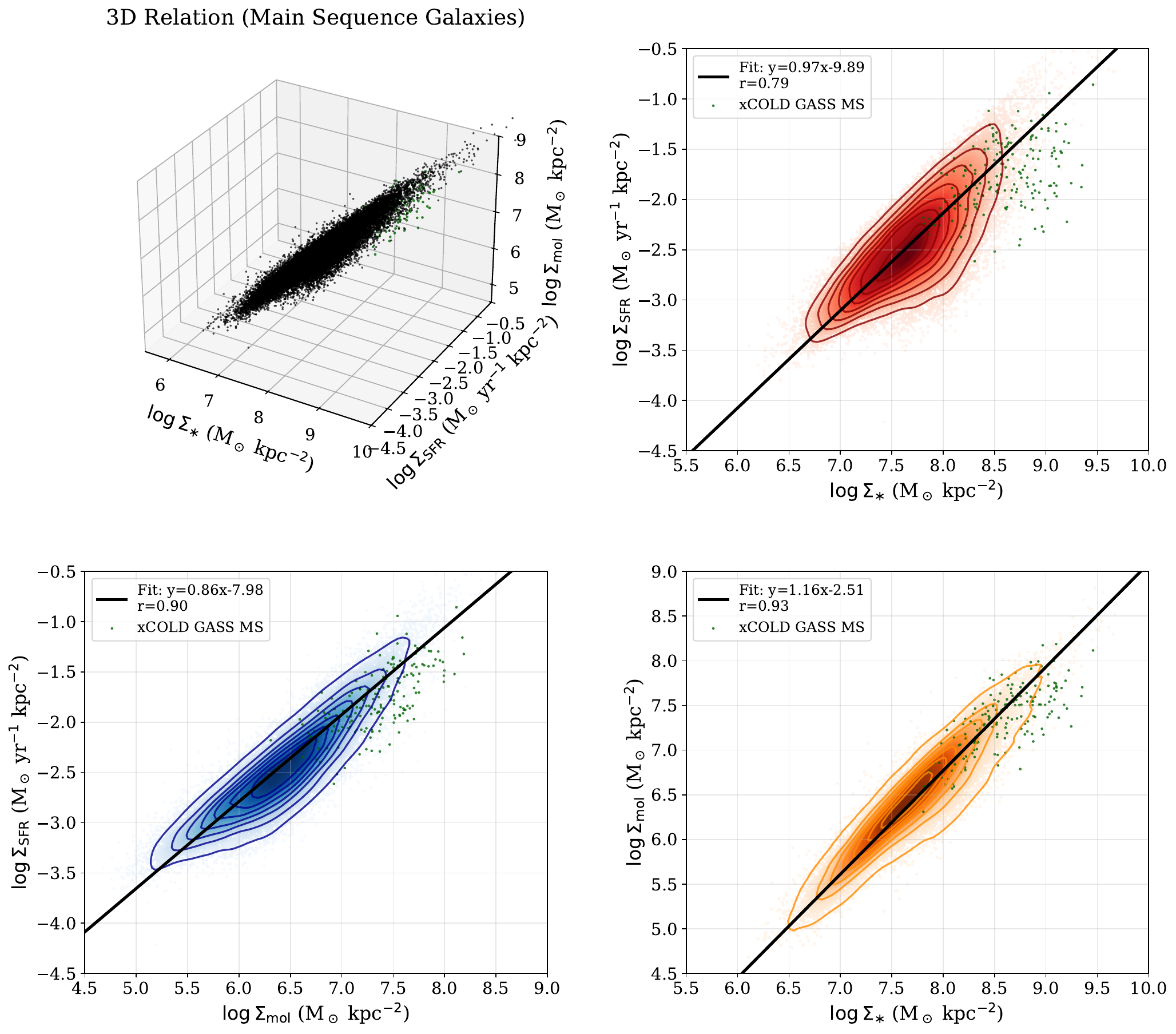}
\caption{The global scaling relations between the star formation rate surface density ($\Sigma_{\rm SFR}$), the stellar mass surface density ($\Sigma_{\ast}$), and the molecular gas mass surface density ($\Sigma_{\rm mol}$) measured at galaxy-wide scales for 24,954  main-sequence galaxies. The top-left panel shows the three-parameter distribution in logarithmic space. The remaining panels present the regression fits to the three projected two-dimensional relations: $\Sigma_{\rm SFR}$ vs. $\Sigma_{\ast}$, $\Sigma_{\rm SFR}$ vs.$\Sigma_{\rm mol}$ ( KS relation), and $\Sigma_{\rm mol}$ vs. $\Sigma_{\ast}$ (MGMS relation). Contours show the distribution of our sample. The best-fit equations, the standard deviation of the residuals ($\sigma$), and the Pearson correlation coefficient (r) are indicated in each panel. The green circles are 198 xCOLD GASS MS galaxies  for demonstration and were not included in the fitting.}\label{fig3}
\end{figure}

Figure~\ref{fig3} presents the scaling relations between the galaxy-wide average surface densities $\Sigma_{\rm SFR}$, $\Sigma_{\ast}$, and $\Sigma_{\rm mol}$ for all main-sequence galaxies in our SDSS volume-limited sample. The three-dimensional distribution in the top-left panel demonstrates a tight, roughly linear correlation between the three parameters in log-space. 

 Through a systematic comparison of our global relations with the resolved (kpc-scale) SFMS, KS, and MGMS relations of \citet{Lin2019}, we establish that they follow similar trends.
 To robustly characterize these relationships, we employed multiple fitting approaches---orthogonal distance regression (ODR), a maximum likelihood (ML)–based method, and a Kolmogorov–Smirnov (KS) test–based method---using the “raddest” package (version 0.9.0) \citep{Jing2024} to fit the two-dimensional projections. Although the ODR slopes (0.97, 0.86, and 1.16) are slightly steeper than those from ML (0.86, 0.86, 1.06) and KS (0.87, 0.82, 1.08), the results from all three methods are consistent and support the subsequent discussion. To facilitate direct comparison with the local study of \citet{Lin2019}, which also used ODR, we report only the ODR best-fit parameters and associated scatters, which are listed in the legends of Figure~\ref{fig3} and in Table~\ref{tab1}. The uncertainties for all three galaxy-wide average surfaces are very small (<0.2 dex) and have no significant impact on the fit.
\begin{table}[htbp]
\scriptsize
\caption{Best-fit parameters for the 2D scaling relations.\label{tab1}}
\begin{tabularx}{\linewidth}{lcccc}
\toprule
\textbf{Relation}	& \textbf{Slope $k$}	& \textbf{Intercept $b$} & \textbf{Correlation coefficient $r$} & \textbf{Scatter $\sigma$ ($\sigma_{\rm int}$)}\\ 
\midrule
$\Sigma_{\rm SFR}$ vs. $\Sigma_{\ast}$		& 0.97 $\pm$ 0.01	& -9.89 $\pm$ 0.03	& 0.79	& 0.30 (0.27) dex \\
$\Sigma_{\rm SFR}$ vs. $\Sigma_{\rm mol}$		& 0.86 $\pm$ 0.01	& -7.89 $\pm$ 0.02	& 0.90  	& 0.22 (0.18) dex \\
$\Sigma_{\rm mol}$ vs. $\Sigma_{\ast}$		& 1.16 $\pm$ 0.01	& -2.51 $\pm$ 0.02	& 0.93	& 0.20 (0.18) dex \\
\bottomrule
\end{tabularx}
\end{table}

The global relation between $\Sigma_{\rm SFR}$ and $\Sigma_{\ast}$, corresponding to the local/resolved SFMS, is observed with a slope of 0.97 $\pm$ 0.01. 
Notably, this relation exhibited the largest scatter ($\sigma$ = 0.30 dex) and the lowest correlation coefficient ($r$ = 0.79) between the three relations.  
 The strong Pearson correlation ($r$ = 0.90) and relatively small scatter ($\sigma$ = 0.22 dex) underscore the direct link between molecular gas mass and star formation activity.
Regarding the MGMS, we identified the strongest correlation with minimal scatter ($\sigma$ = 0.20 dex) and the highest correlation coefficient ($r$ = 0.93). 
When we analyze the 198 main-sequence galaxies from the xCOLD GASS sample with direct CO detections, the ordering of correlation coefficients and scatter among these three relations remains the same as in our main sample, although the best-fit parameters are subject to relatively large uncertainties. These results indicate that the distributions of molecular gas and stellar mass are closely coupled \citep{Wong2013ApJ...777L...4W}, and thus are potentially governed by the same gravitational potential. 

\subsection{Disentangling the Primary Drivers: Results from Multivariate Regression and Partial Correlation Analysis}\label{sub_result3}

While the comparison of scatter and correlations coefficient  in the 2D projections (Section~\ref{sub_result2}) provides strong evidence for the primacy of the gas relations, a more rigorous test is to statistically control for confounding variables. We therefore performed multivariate regression and partial correlation analysis on our dataset of 24,954 galaxies. 

The regression yields the following relations: $\log\Sigma_{\rm SFR}= -0.3\log\Sigma_{\ast} +1.05 \log\Sigma_{\rm mol} -6.88$, and 
$\log\Sigma_{\rm mol} = 0.69\log\Sigma_{\ast} + 0.47\log\Sigma_{\rm SFR} + 2.31$. 
The coefficients from these regressions indicate the independent contribution of each predictor. 
The multivariate regression reveals that when  both $\Sigma_{\ast}$ and $\Sigma_{\rm mol}$ are considered simultaneously, the coefficient for $\log\Sigma_{\ast}$ becomes negative (-0.3). 
This negative dependence of $\Sigma_{\rm SFR}$ on $\Sigma_{\ast}$ at fixed $\Sigma_{\rm mol}$ is remarkably consistent with previous studies (e.g., \citet{Lin2019,Jing2024b}).
In contrast, the coefficient for $\log\Sigma_{\rm mol}$ remains strongly positive (1.05), highlighting its dominant role.

This finding is powerfully corroborated by the partial correlation analysis. After controlling for the molecular gas surface density, the partial correlation between $\Sigma_{\rm SFR}$ and $\Sigma_{\ast}$ vanishes and becomes slightly negative ($r$=-0.24). Conversely, the partial correlation between $\Sigma_{\rm SFR}$ and $\Sigma_{\rm mol}$, while controlling for $\Sigma_{\ast}$, remains strong and highly significant ($r$=0.71). The strongest partial correlation($r$=0.81) is found between $\Sigma_{\rm mol}$ and $\Sigma_{\ast}$ after controlling for $\Sigma_{\rm SFR}$,
reinforcing the tight coupling between a galaxy's stellar mass and its molecular gas reservoir.

\section{Discussion}
\label{sec:Discussion}
\subsection{The Connection between Local and Global Relations}\label{sec:Discuss1}
While derived at different spatial resolutions, our galaxy-wide average densities probe a similar—and, in part, broader—parameter range than the resolved measurements of \citet{Lin2019}. Notably, 
the trends of our global relations align with their local counterparts---a central finding of this study.
 
We employed ODR to fit the two-dimensional scaling relations. This choice was motivated by two primary considerations. First, ODR symmetrically accounts for the uncertainties in both variables, which is appropriate given the comparable magnitude of observational errors in our derived surface densities \citep{2021MNRAS.503.1615S}. Second, it ensures a direct and consistent comparison with the work of \citet{Lin2019}, who applied the same methodology to resolved data. While alternative techniques such as principal component analysis could describe the multi-dimensional parameter space, ODR provides a straightforward and interpretable fit for each projected relation, which is sufficient for our goal of comparing slopes and scatters across studies and avoids potential biases from differing fitting procedures.

Figure~\ref{fig3} and Table~\ref{tab1} shows that when the star formation density is the $y$-axis, our slopes show slight variations compared with those of \citet{Lin2019} (e.g., 0.97 to 1.19 for $\Sigma_{\ast}$ vs. $\Sigma_{\rm SFR}$, 0.86 to 1.05 for $\Sigma_{\rm mol}$ vs. $\Sigma_{\rm SFR}$).  
Considering the similar results for $\Sigma_{\rm mol}$ vs. $\Sigma_{\ast}$, this kind of shallower slope in global relations should be real; although the slope of the resolved SFMS and KS relationship also varies \citep{Hsieh2017,Bolatto2017,Ellison2020MNRAS.493L..39E}. These discrepancies are likely attributable to differences in sample selection (see Subsection~\ref{sub_result1}) and the methodological details involved in computing surface densities (outlined in Subsection~\ref{Properties}). A key factor is that our linear regression was performed on galaxy-integrated quantities, whereas \citet{Lin2019} conducted their analysis using only star-forming spaxels. The inclusion of non-star-forming regions in integrated measurements is known to systematically lower the derived slopes---an effect consistent with findings from \citet{Cano2016} and \citet{Cano2019,2021MNRAS.503.1615S}. 

\subsection{The Relation between $\Sigma_{\rm SFR}$ and $\Sigma_{\ast}$ can be Explained by the Combination of the SK and MGMS }
Despite these quantitative differences, the relative ordering of scatter and correlation strength between the three relations is also consistent with the resolved analyses. Therefore, the primary conclusion of our analysis remains robust: the global relation between $\Sigma_{\rm SFR}$ and  $\Sigma_{\ast}$ exhibits the largest scatter and the weakest correlation between the three scaling relations. 

This result can be understood within a framework where the star formation rate is primarily set by the molecular gas reservoir. If the star formation surface density in a galaxy is governed by the molecular gas surface density, $\log\Sigma_{\rm SFR} = k_1 \log\Sigma_{\rm mol}+{c_1}$, and the molecular gas is coupled to the stellar component, $\log\Sigma_{\rm mol} = k_2 \log\Sigma_{\ast}+{c_2}$, then the relation between $\Sigma_{\rm SFR}$ and  $\Sigma_{\ast}$ emerges as a direct consequence:
$\Sigma_{\rm SFR} = k_1 k_2 \log\Sigma_{\ast}+(c_1+ k_1 \times c_2)$.
The derived formula is $\textrm{log}\Sigma_{\rm SFR} =1.0 \textrm{log}\Sigma_{\ast} - 10.13$, which closely matches our fitted one: ($\textrm{log}\Sigma_{\rm SFR} =0.97 \textrm{log}\Sigma_{\ast} - 9.89$).
Furthermore, the observed scatter in this global relation is consistent with the expected dispersion for the combination of the other two relations, calculated as the quadratic sum of their individual scatters ($\sqrt{0.22^2 + 0.20^2} \approx 0.30$ dex).

Therefore, our results provide support for the scenario proposed in \citet{Lin2019,Ellison2020MNRAS.493L..39E,Baker2022MNRAS.510.3622B}, wherein the global relation between $\Sigma_{\rm SFR}$ and  $\Sigma_{\ast}$ is not a fundamental physical relation but, rather, emerges from the combination of the more fundamental KS and MGMS relations. 

It is worth noting that our analyses were based on galaxy-integrated quantities; thus, these results statistically validate the above scenario at global, galaxy-wide scales. Compared with resolved MGMS, the global MGMS relation between $\Sigma_{\rm mol}$ and $\Sigma_{\ast}$ can be easily understood through a straightforward physical picture: gas accretion and retention on galactic scales are governed by the depth of the gravitational potential well \citep{Barrera2018,Micha2024} and the transition from atomic gas under pressure \citep{Somerville2015}, which are linked to the total stellar mass of the galaxy. 
Our study provides statistical validation of this picture at galaxy-wide scales, using a sample an order of magnitude larger than those used in previous CO-based works.

The multivariate regression and partial correlation analysis in Section~\ref{sub_result3} place this conclusion on an even firmer statistical footing. By directly controlling for $\Sigma_{\rm mol}$, we demonstrate that the residual correlation between $\Sigma_{\rm SFR}$ and $\Sigma_{\ast}$ is effectively eliminated, confirming that stellar mass plays no direct role in setting the SFR once the gas content is accounted for. 

The role of $\Sigma_{\ast}$ in regulating star formation appears to be complex. While our work—along with \citet{Lin2019}—finds a negative residual correlation between $\Sigma_{\rm SFR}$ and $\Sigma_{\ast}$ after controlling for $\Sigma_{\rm mol}$, other studies have reported a positive effect (e.g., \citet{Shi2011,Shi2018}). As discussed by \citet{Jing2024b}, this discrepancy likely stems from differences in the parameter space probed: the positive correlation tends to emerge in gas-dominated, low-surface-brightness regions where the gravitational potential of existing stars may facilitate gas collapse, whereas the negative correlation is observed in star-dominated regions similar to our sample. Furthermore, \citet{Pessa2022} demonstrated that the measured power-law exponent for $\Sigma_{\ast}$ can vary significantly with the spatial scale of analysis. Our results, obtained at galaxy-wide integrated scales, are consistent with the negative exponent found at kpc scales \citep{Lin2019}, reinforcing the notion that in high-surface-density environments, the primary role of stellar mass is not to directly fuel star formation but to trace the molecular gas reservoir.

Looking forward, the next step will be to investigate the spatial scale dependence of these relations; for example, utilizing resolved CO maps from direct observations or estimated distributions \citep{Gao2022}.

\section{Conclusions}
\label{sec:summary}
In this study, we derived global relations involving the galaxy-wide averages of  $\Sigma_{\rm SFR}$, $\Sigma_{\ast}$, and $\Sigma_{\rm mol}$ based on a statistically constructed, volume-limited sample of 24,954 main-sequence galaxies, then compared them with local/resolved relations between the same quantities. Our main conclusions are summarized as follows:
\begin{enumerate}
\item	We determined that the galaxy-wide averages of $\Sigma_{\rm SFR}$, $\Sigma_{\ast}$, and $\Sigma_{\rm mol}$ correlate with one another, and the correlations are similar to kpc-scale local/resolved relations.
\item	Compared with KS and MGMS relations, the relation between $\Sigma_{\rm SFR}$ and  $\Sigma_{\ast}$ showed the largest scatter and weakest correlation, suggesting that it may be a secondary consequence of the two other fundamental relations.
\item	Multivariate regression and partial correlation analyses reveal that once $\Sigma_{\rm mol}$ is controlled, the correlation between $\Sigma_{\rm SFR}$ and  $\Sigma_{\ast}$
disappears, with the residual dependence becoming slightly negative.
\item	The best-fitted linear relation between $\Sigma_{\rm SFR}$ and  $\Sigma_{\ast}$ can be precisely predicted mathematically from the established KS and MGMS relations, providing direct quantitative support for its emergent nature.
\end{enumerate}

Collectively, our results favor a gas-driven paradigm in which star formation is directly fueled by molecular gas, where gas content is coupled to the stellar gravitational potential at both galaxy-wide and local scales, and stellar mass plays no direct role in setting the star formation rate at fixed gas content.

\vspace{6pt} 

\section*{Acknowledgments}
We thank the anonymous referees for a thorough and helpful report.


\begin{thebibliography}{}
\expandafter\ifx\csname natexlab\endcsname\relax\def\natexlab#1{#1}\fi
\providecommand{\url}[1]{\href{#1}{#1}}

\bibitem[{{Baker} {et~al.}(2022){Baker}, {Maiolino}, {Bluck}, {Lin}, {Ellison}, {Belfiore}, {Pan}, \& {Thorp}}]{Baker2022MNRAS.510.3622B}
{Baker}, W.~M., {Maiolino}, R., {Bluck}, A. F.~L., {et~al.} 2022, \mnras, 510, 3622

\bibitem[{{Barrera-Ballesteros} {et~al.}(2018){Barrera-Ballesteros}, {Heckman}, {S{\'a}nchez}, {Zakamska}, {Cleary}, {Zhu}, {Brinkmann}, {Drory}, \& {THE MaNGA TEAM}}]{Barrera2018}
{Barrera-Ballesteros}, J.~K., {Heckman}, T., {S{\'a}nchez}, S.~F., {et~al.} 2018, \apj, 852, 74

\bibitem[{{Barrera-Ballesteros} {et~al.}(2021){Barrera-Ballesteros}, {Heckman}, {S{\'a}nchez}, {Drory}, {Cruz-Gonzalez}, {Carigi}, {Riffel}, {Boquien}, {Tissera}, {Bizyaev}, {Rong}, {Boardman}, {Alvarez Hurtado}, \& {MaNGA Team}}]{2021ApJ...909..131B}
---. 2021, \apj, 909, 131

\bibitem[{{Blanton} {et~al.}(2011){Blanton}, {Kazin}, {Muna}, {Weaver}, \& {Price-Whelan}}]{Blanton2011}
{Blanton}, M.~R., {Kazin}, E., {Muna}, D., {Weaver}, B.~A., \& {Price-Whelan}, A. 2011, \aj, 142, 31

\bibitem[{{Bolatto} {et~al.}(2017){Bolatto}, {Wong}, {Utomo}, {Blitz}, {Vogel}, {S{\'a}nchez}, {Barrera-Ballesteros}, {Cao}, {Colombo}, {Dannerbauer}, {Garc{\'\i}a-Benito}, {Herrera-Camus}, {Husemann}, {Kalinova}, {Leroy}, {Leung}, {Levy}, {Mast}, {Ostriker}, {Rosolowsky}, {Sandstrom}, {Teuben}, {van de Ven}, \& {Walter}}]{Bolatto2017}
{Bolatto}, A.~D., {Wong}, T., {Utomo}, D., {et~al.} 2017, \apj, 846, 159

\bibitem[{{Brinchmann} {et~al.}(2004){Brinchmann}, {Charlot}, {White}, {Tremonti}, {Kauffmann}, {Heckman}, \& {Brinkmann}}]{Brinchmann2004}
{Brinchmann}, J., {Charlot}, S., {White}, S.~D.~M., {et~al.} 2004, \mnras, 351, 1151

\bibitem[{{Cano-D{\'\i}az} {et~al.}(2019){Cano-D{\'\i}az}, {{\'A}vila-Reese}, {S{\'a}nchez}, {Hern{\'a}ndez-Toledo}, {Rodr{\'\i}guez-Puebla}, {Boquien}, \& {Ibarra-Medel}}]{Cano2019}
{Cano-D{\'\i}az}, M., {{\'A}vila-Reese}, V., {S{\'a}nchez}, S.~F., {et~al.} 2019, \mnras, 488, 3929

\bibitem[{{Cano-D{\'\i}az} {et~al.}(2016){Cano-D{\'\i}az}, {S{\'a}nchez}, {Zibetti}, {Ascasibar}, {Bland-Hawthorn}, {Ziegler}, {Gonz{\'a}lez Delgado}, {Walcher}, {Garc{\'\i}a-Benito}, {Mast}, {Mendoza-P{\'e}rez}, {Falc{\'o}n-Barroso}, {Galbany}, {Husemann}, {Kehrig}, {Marino}, {S{\'a}nchez-Bl{\'a}zquez}, {L{\'o}pez-Cob{\'a}}, {L{\'o}pez-S{\'a}nchez}, \& {Vilchez}}]{Cano2016}
{Cano-D{\'\i}az}, M., {S{\'a}nchez}, S.~F., {Zibetti}, S., {et~al.} 2016, \apjl, 821, L26

\bibitem[{{Daddi} {et~al.}(2007){Daddi}, {Dickinson}, {Morrison}, {Chary}, {Cimatti}, {Elbaz}, {Frayer}, {Renzini}, {Pope}, {Alexander}, {Bauer}, {Giavalisco}, {Huynh}, {Kurk}, \& {Mignoli}}]{2007ApJ...670..156D}
{Daddi}, E., {Dickinson}, M., {Morrison}, G., {et~al.} 2007, \apj, 670, 156

\bibitem[{{Ellison} {et~al.}(2020){Ellison}, {Thorp}, {Lin}, {Pan}, {Bluck}, {Scudder}, {Teimoorinia}, {S{\'a}nchez}, \& {Sargent}}]{Ellison2020MNRAS.493L..39E}
{Ellison}, S.~L., {Thorp}, M.~D., {Lin}, L., {et~al.} 2020, \mnras, 493, L39

\bibitem[{{Gao} {et~al.}(2022){Gao}, {Tan}, {Gao}, {Fang}, {Chown}, {Jiao}, \& {Luo}}]{Gao2022}
{Gao}, Y., {Tan}, Q.-H., {Gao}, Y., {et~al.} 2022, \apj, 940, 133

\bibitem[{{Gao} {et~al.}(2019){Gao}, {Xiao}, {Li}, {Jiang}, {Tan}, {Gao}, {Wilson}, {Bureau}, {Saintonge}, {S{\'a}nchez-Gallego}, {Brown}, {Clark}, {Hwang}, {Lamperti}, {Lin}, {Liu}, {Lu}, {Pan}, {Sun}, \& {Williams}}]{Gao2019}
{Gao}, Y., {Xiao}, T., {Li}, C., {et~al.} 2019, \apj, 887, 172

\bibitem[{{Gao} {et~al.}(2025){Gao}, {Wang}, {Tan}, {Davis}, {Liang}, {Jiang}, {Gai}, {Jiao}, {Shi}, {Feng}, {Tang}, {Li}, \& {Wang}}]{Gao2025}
{Gao}, Y., {Wang}, E., {Tan}, Q.-H., {et~al.} 2025, \apj, 979, 105

\bibitem[{{Hsieh} {et~al.}(2017){Hsieh}, {Lin}, {Lin}, {Pan}, {Hsu}, {S{\'a}nchez}, {Cano-D{\'\i}az}, {Zhang}, {Yan}, {Barrera-Ballesteros}, {Boquien}, {Riffel}, {Brownstein}, {Cruz-Gonz{\'a}lez}, {Hagen}, {Ibarra}, {Pan}, {Bizyaev}, {Oravetz}, \& {Simmons}}]{Hsieh2017}
{Hsieh}, B.~C., {Lin}, L., {Lin}, J.~H., {et~al.} 2017, \apjl, 851, L24

\bibitem[{{Jarrett} {et~al.}(2011){Jarrett}, {Cohen}, {Masci}, {Wright}, {Stern}, {Benford}, {Blain}, {Carey}, {Cutri}, {Eisenhardt}, {Lonsdale}, {Mainzer}, {Marsh}, {Padgett}, {Petty}, {Ressler}, {Skrutskie}, {Stanford}, {Surace}, {Tsai}, {Wheelock}, \& {Yan}}]{Jarrett2011}
{Jarrett}, T.~H., {Cohen}, M., {Masci}, F., {et~al.} 2011, The Astrophysical Journal, 735, 112

\bibitem[{{Jing} \& {Li}(2024{\natexlab{a}})}]{Jing2024}
{Jing}, T., \& {Li}, C. 2024{\natexlab{a}}, arXiv e-prints, arXiv:2411.08747

\bibitem[{{Jing} \& {Li}(2024{\natexlab{b}})}]{Jing2024b}
---. 2024{\natexlab{b}}, \apj, 975, 17

\bibitem[{{Leroy} {et~al.}(2008){Leroy}, {Walter}, {Brinks}, {Bigiel}, {de Blok}, {Madore}, \& {Thornley}}]{Leroy2008}
{Leroy}, A.~K., {Walter}, F., {Brinks}, E., {et~al.} 2008, \aj, 136, 2782

\bibitem[{{Leroy} {et~al.}(2009){Leroy}, {Walter}, {Bigiel}, {Usero}, {Weiss}, {Brinks}, {de Blok}, {Kennicutt}, {Schuster}, {Kramer}, {Wiesemeyer}, \& {Roussel}}]{Leroy2009}
{Leroy}, A.~K., {Walter}, F., {Bigiel}, F., {et~al.} 2009, \aj, 137, 4670

\bibitem[{{Lilly} {et~al.}(2013){Lilly}, {Carollo}, {Pipino}, {Renzini}, \& {Peng}}]{2013ApJ...772..119L}
{Lilly}, S.~J., {Carollo}, C.~M., {Pipino}, A., {Renzini}, A., \& {Peng}, Y. 2013, \apj, 772, 119

\bibitem[{{Lin} {et~al.}(2019){Lin}, {Pan}, {Ellison}, {Belfiore}, {Shi}, {S{\'a}nchez}, {Hsieh}, {Rowlands}, {Ramya}, {Thorp}, {Li}, \& {Maiolino}}]{Lin2019}
{Lin}, L., {Pan}, H.-A., {Ellison}, S.~L., {et~al.} 2019, \apjl, 884, L33

\bibitem[{{Micha{\l}owski} {et~al.}(2024){Micha{\l}owski}, {Gall}, {Hjorth}, {Frayer}, {Tsai}, {Rowlands}, {Takeuchi}, {Le{\'s}niewska}, {Behrendt}, {Bourne}, {Hughes}, {Koprowski}, {Nadolny}, {Ryzhov}, {Solar}, {Spring}, {Zavala}, \& {Bartczak}}]{Micha2024}
{Micha{\l}owski}, M.~J., {Gall}, C., {Hjorth}, J., {et~al.} 2024, \apj, 964, 129

\bibitem[{{Ostriker} {et~al.}(2010){Ostriker}, {McKee}, \& {Leroy}}]{Ostriker2010}
{Ostriker}, E.~C., {McKee}, C.~F., \& {Leroy}, A.~K. 2010, \apj, 721, 975

\bibitem[{{Pessa} {et~al.}(2022){Pessa}, {Schinnerer}, {Leroy}, {Koch}, {Rosolowsky}, {Williams}, {Pan}, {Schruba}, {Usero}, {Belfiore}, {Bigiel}, {Blanc}, {Chevance}, {Dale}, {Emsellem}, {Gensior}, {Glover}, {Grasha}, {Groves}, {Klessen}, {Kreckel}, {Kruijssen}, {Liu}, {Meidt}, {Pety}, {Querejeta}, {Saito}, {Sanchez-Blazquez}, \& {Watkins}}]{Pessa2022}
{Pessa}, I., {Schinnerer}, E., {Leroy}, A.~K., {et~al.} 2022, \aap, 663, A61

\bibitem[{{Saintonge} {et~al.}(2011){Saintonge}, {Kauffmann}, {Wang}, {Kramer}, {Tacconi}, {Buchbender}, {Catinella}, {Graci{\'a}-Carpio}, {Cortese}, {Fabello}, {Fu}, {Genzel}, {Giovanelli}, {Guo}, {Haynes}, {Heckman}, {Krumholz}, {Lemonias}, {Li}, {Moran}, {Rodriguez-Fernandez}, {Schiminovich}, {Schuster}, \& {Sievers}}]{Saintonge2011b}
{Saintonge}, A., {Kauffmann}, G., {Wang}, J., {et~al.} 2011, \mnras, 415, 61

\bibitem[{{Saintonge} {et~al.}(2012){Saintonge}, {Tacconi}, {Fabello}, {Wang}, {Catinella}, {Genzel}, {Graci{\'a}-Carpio}, {Kramer}, {Moran}, {Heckman}, {Schiminovich}, {Schuster}, \& {Wuyts}}]{Saintonge2012}
{Saintonge}, A., {Tacconi}, L.~J., {Fabello}, S., {et~al.} 2012, \apj, 758, 73

\bibitem[{{Saintonge} {et~al.}(2016){Saintonge}, {Catinella}, {Cortese}, {Genzel}, {Giovanelli}, {Haynes}, {Janowiecki}, {Kramer}, {Lutz}, {Schiminovich}, {Tacconi}, {Wuyts}, \& {Accurso}}]{Saintonge2016}
{Saintonge}, A., {Catinella}, B., {Cortese}, L., {et~al.} 2016, \mnras, 462, 1749

\bibitem[{{Saintonge} {et~al.}(2017){Saintonge}, {Catinella}, {Tacconi}, {Kauffmann}, {Genzel}, {Cortese}, {Dav{\'e}}, {Fletcher}, {Graci{\'a}-Carpio}, {Kramer}, {Heckman}, {Janowiecki}, {Lutz}, {Rosario}, {Schiminovich}, {Schuster}, {Wang}, {Wuyts}, {Borthakur}, {Lamperti}, \& {Roberts-Borsani}}]{Saintonge2017}
{Saintonge}, A., {Catinella}, B., {Tacconi}, L.~J., {et~al.} 2017, \apjs, 233, 22

\bibitem[{{Salim} {et~al.}(2018){Salim}, {Boquien}, \& {Lee}}]{Salim2018}
{Salim}, S., {Boquien}, M., \& {Lee}, J.~C. 2018, \apj, 859, 11

\bibitem[{{Salim} {et~al.}(2007){Salim}, {Rich}, {Charlot}, {Brinchmann}, {Johnson}, {Schiminovich}, {Seibert}, {Mallery}, {Heckman}, {Forster}, {Friedman}, {Martin}, {Morrissey}, {Neff}, {Small}, {Wyder}, {Bianchi}, {Donas}, {Lee}, {Madore}, {Milliard}, {Szalay}, {Welsh}, \& {Yi}}]{2007ApJS..173..267S}
{Salim}, S., {Rich}, R.~M., {Charlot}, S., {et~al.} 2007, \apjs, 173, 267

\bibitem[{{Salim} {et~al.}(2016){Salim}, {Lee}, {Janowiecki}, {da Cunha}, {Dickinson}, {Boquien}, {Burgarella}, {Salzer}, \& {Charlot}}]{Salim2016}
{Salim}, S., {Lee}, J.~C., {Janowiecki}, S., {et~al.} 2016, \apjs, 227, 2

\bibitem[{{S{\'a}nchez}(2020)}]{Sanchez2020}
{S{\'a}nchez}, S.~F. 2020, \araa, 58, 99

\bibitem[{{S{\'a}nchez} {et~al.}(2021){S{\'a}nchez}, {Barrera-Ballesteros}, {Colombo}, {Wong}, {Bolatto}, {Rosolowsky}, {Vogel}, {Levy}, {Kalinova}, {Alvarez-Hurtado}, {Luo}, \& {Cao}}]{2021MNRAS.503.1615S}
{S{\'a}nchez}, S.~F., {Barrera-Ballesteros}, J.~K., {Colombo}, D., {et~al.} 2021, \mnras, 503, 1615

\bibitem[{{Schinnerer} \& {Leroy}(2024)}]{Schinnerer2024}
{Schinnerer}, E., \& {Leroy}, A.~K. 2024, \araa, 62, 369

\bibitem[{{Shi} {et~al.}(2011){Shi}, {Helou}, {Yan}, {Armus}, {Wu}, {Papovich}, \& {Stierwalt}}]{Shi2011}
{Shi}, Y., {Helou}, G., {Yan}, L., {et~al.} 2011, \apj, 733, 87

\bibitem[{{Shi} {et~al.}(2018){Shi}, {Yan}, {Armus}, {Gu}, {Helou}, {Qiu}, {Gwyn}, {Stierwalt}, {Fang}, {Chen}, {Zhou}, {Wu}, {Zheng}, {Zhang}, {Gao}, \& {Wang}}]{Shi2018}
{Shi}, Y., {Yan}, L., {Armus}, L., {et~al.} 2018, \apj, 853, 149

\bibitem[{{Somerville} \& {Dav{\'e}}(2015)}]{Somerville2015}
{Somerville}, R.~S., \& {Dav{\'e}}, R. 2015, \araa, 53, 51

\bibitem[{{Wang} {et~al.}(2019){Wang}, {Lilly}, {Pezzulli}, \& {Matthee}}]{wang2019}
{Wang}, E., {Lilly}, S.~J., {Pezzulli}, G., \& {Matthee}, J. 2019, \apj, 877, 132

\bibitem[{{Wong} {et~al.}(2013){Wong}, {Xue}, {Bolatto}, {Leroy}, {Blitz}, {Rosolowsky}, {Bigiel}, {Fisher}, {Ott}, {Rahman}, {Vogel}, \& {Walter}}]{Wong2013ApJ...777L...4W}
{Wong}, T., {Xue}, R., {Bolatto}, A.~D., {et~al.} 2013, \apjl, 777, L4

\bibitem[{{Wright} {et~al.}(2010){Wright}, {Eisenhardt}, {Mainzer}, {Ressler}, {Cutri}, {Jarrett}, {Kirkpatrick}, {Padgett}, {McMillan}, {Skrutskie}, {Stanford}, {Cohen}, {Walker}, {Mather}, {Leisawitz}, {Gautier}, {McLean}, {Benford}, {Lonsdale}, {Blain}, {Mendez}, {Irace}, {Duval}, {Liu}, {Royer}, {Heinrichsen}, {Howard}, {Shannon}, {Kendall}, {Walsh}, {Larsen}, {Cardon}, {Schick}, {Schwalm}, {Abid}, {Fabinsky}, {Naes}, \& {Tsai}}]{Wright2010}
{Wright}, E.~L., {Eisenhardt}, P. R.~M., {Mainzer}, A.~K., {et~al.} 2010, \aj, 140, 1868

\bibitem[{{Zheng} {et~al.}(2013){Zheng}, {Meurer}, {Heckman}, {Thilker}, \& {Zwaan}}]{Zheng2013}
{Zheng}, Z., {Meurer}, G.~R., {Heckman}, T.~M., {Thilker}, D.~A., \& {Zwaan}, M.~A. 2013, \mnras, 434, 3389

\end{thebibliography}
\end{document}